\documentclass[aps,twocolumn,showpacs]{revtex4}
\usepackage{graphicx}
\usepackage{epsf}
\usepackage{amsmath}

\usepackage{amssymb}
\begin{document}
%TIolo
\title{Shock waves in disordered media}
\vspace{0.5cm}
%Autori
\author{N. Ghofraniha$^{1\star}$, S. Gentilini$^2$, V. Folli$^2$, E. DelRe$^{3,1}$ and C. Conti$^{3,2}$}
\affiliation{\small
$^1$ IPCF-CNR c/o Dipartimento di Fisica - Universit\`{a} La Sapienza, P. A. Moro
2, 00185, Roma, Italy\\
{$^2$ ISC-CNR c/o Dipartimento di Fisica - Universit\`{a} La Sapienza, P. A. Moro
2, 00185, Roma, Italy\\
$^3$ Dipartimento di Fisica - Universit\`{a} La Sapienza, P. A. Moro
2, 00185, Roma, Italy\\
$^*$Corresponding author: neda.ghofraniha@roma1.infn.it
 } }
 %Data
\date{\today}
%Abstract
\begin{abstract}
We experimentally investigate the interplay between spatial shock waves and the degree of disorder during nonlinear optical propagation in a thermal defocusing medium.
We characterize the way the shock point is affected by the amount of disorder and scales with wave amplitude.
Evidence for the existence of a phase diagram in terms of nonlinearity and amount of randomness is reported.
The results are in quantitative agreement with a theoretical approach based on the hydrodynamic approximation.
\end{abstract}
%PACS

\maketitle
Laser beams propagating in nonlinear media undergo severe distortions as the power is increased: 
spreading due to diffraction can be progressively reduced through self narrowing,  up to the generation of solitons~\cite{TrilloBook,KivsharBook}  and
dissipative and dispersive shock waves (SWs)~\cite{ gurevich, kamchatnov, rothenberg,wan,conti09,con10},
thus fostering the formation of a variety of nonlinear waves.
The way these are affected by disorder  is a leading main-stream of modern 
research~\cite{Conti07, kivshar10,Fol12,Levi11}.
%staliunas03,swartz07, modugno2010,trillo2011,Val11,Fleisher2012,Lahini09,,Jia10}.
Attention is given to the competition between strongly nonlinear and coherent phenomena, and their frustration due to randomness and scattering; 
recent theoretical investigations deal with general frameworks described by ``phase-diagrams" in terms of the two parameters: the amount of nonlinearity and of 
disorder~\cite{Con11}. However, no direct experimental nonlinearity-disorder phase diagram has been reported.

The case of SWs is specifically relevant~\cite{ghofraniha,  barsi:07,Gho12}, as they represent a strongly nonlinear and coherent
oscillation (the {\it undular bore})~ \cite{deykoon, wyller, conti, wurtz,con10} and are expected to be strongly affected (and eventually inhibited) by disorder,
at variance, e.g., with solitons, which can survive a certain amount of randomness (see, e.g.,\cite{kartashov08,Folli10}). This leads to the direct opposition between the two effects: 
on one hand increasing the nonlinearity favors the shock formation, on the other hand disorder-induced scattering limits this phenomenon. This is relevant in colloidal 
systems~\cite{Conti05PRL, Con06, Dholakia07, Matuszewski09, Lee09, con10}
where disorder is unavoidable, as well as in out-of-equilibrium photorefractive nonlinearities~\cite{DelRe11}, optical fibers~\cite{rothenberg,conti10},
and also in Bose-Einstein condensation~\cite{hoefer,meppelink,dutton} and acoustics~\cite{taylor}.

\begin{figure}[h]
\includegraphics[width=8.5cm]{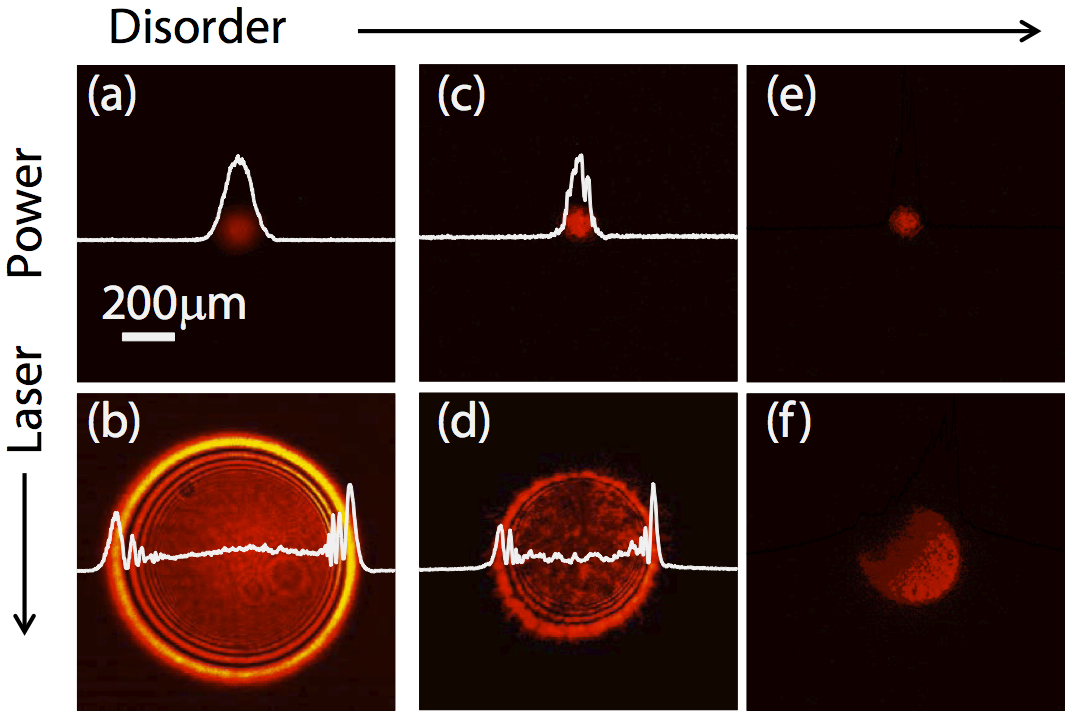}
 \caption{ (Color online)
 Images of transmitted intensity for  different input  power $P$ and   particle concentration $c$:
a) $P$=5 mW, $c$= 0, b) $P$=400 mW, $c$= 0, c) $P$=5 mW, $c$=0.017~w/w, d) $P$=400 mW, $c$=~0.017 w/w e) $P$=5 mW, $c$=~0.030 w/w, f) $P$=400 mW, $c$=0.030~w/w.
Superimposed curves show the measured section of the intensity profiles.\label{figexp1}}
\end{figure}

\begin{figure}[ht]
\includegraphics[width=8.5cm]{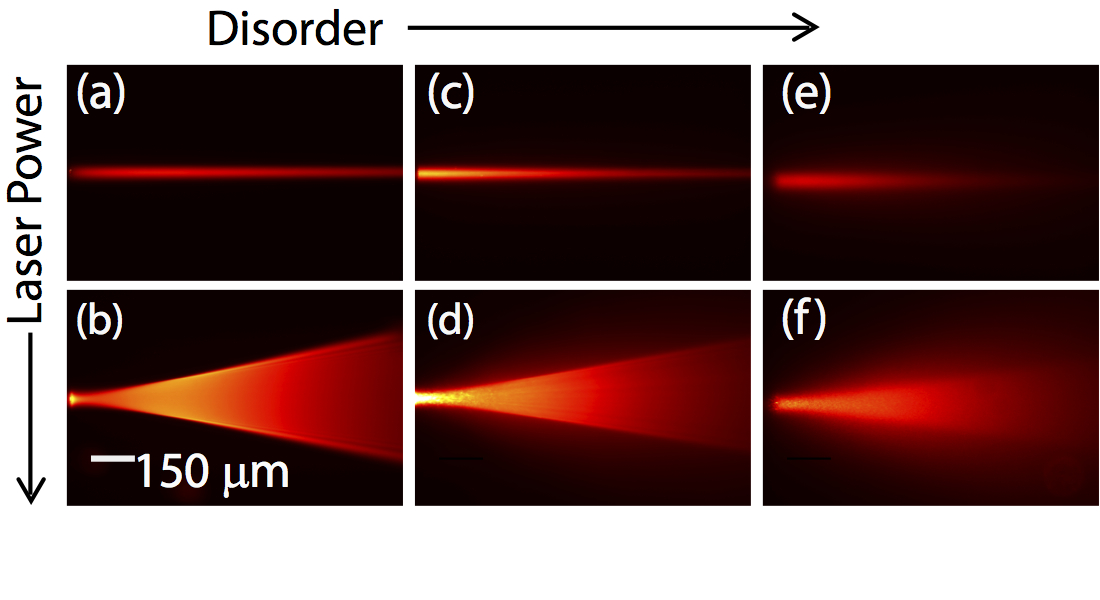}
\caption{(Color online) Beam propagation as observed from top fluorescence emission for different  input power $P$ and particle concentration $c$:
a) $P$=8 mW, $c$= 0, b) $P$=450 mW, $c$= 0, c) $P$=8 mW, $c$=0.017~w/w, d) $P$=450 mW, $c$=0.017~w/w e) $P$=8 mW, $c$=0.030~w/w, f) $P$=450 mW, $c$=0.030~w/w.
\label{figexp2}}
\end{figure}

\begin{figure}[h!]
\includegraphics[width=8.4cm]{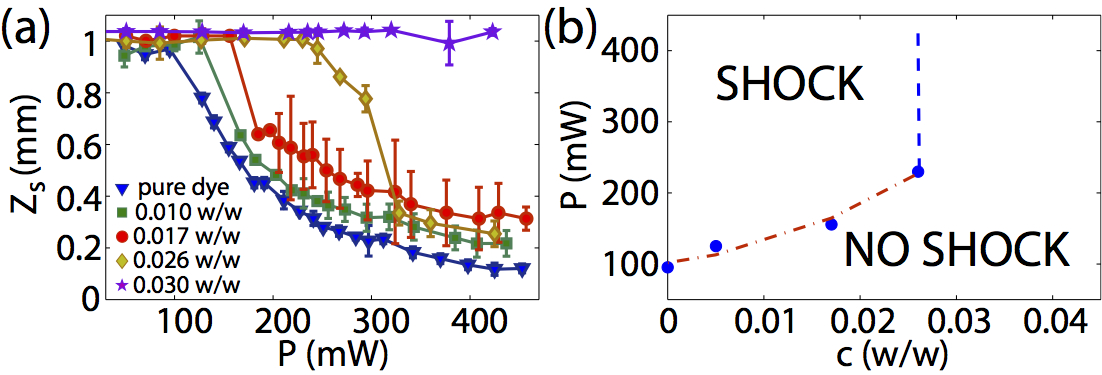}
 \caption{(Color online) (a) Measured shock point $Z_{\textrm{S}}$ Vs $P$ for various $c$;
(b) disorder-power phase diagram with shock and non-shock regimes  obtained from the data in panel (a):
 dots correspond to the threshold powers,  dashed line and the dot-dashed line are the boundaries as estimated by the theory.
\label{figexp3}}
\end{figure}

\begin{figure}[h]
\includegraphics[width=7.5cm]{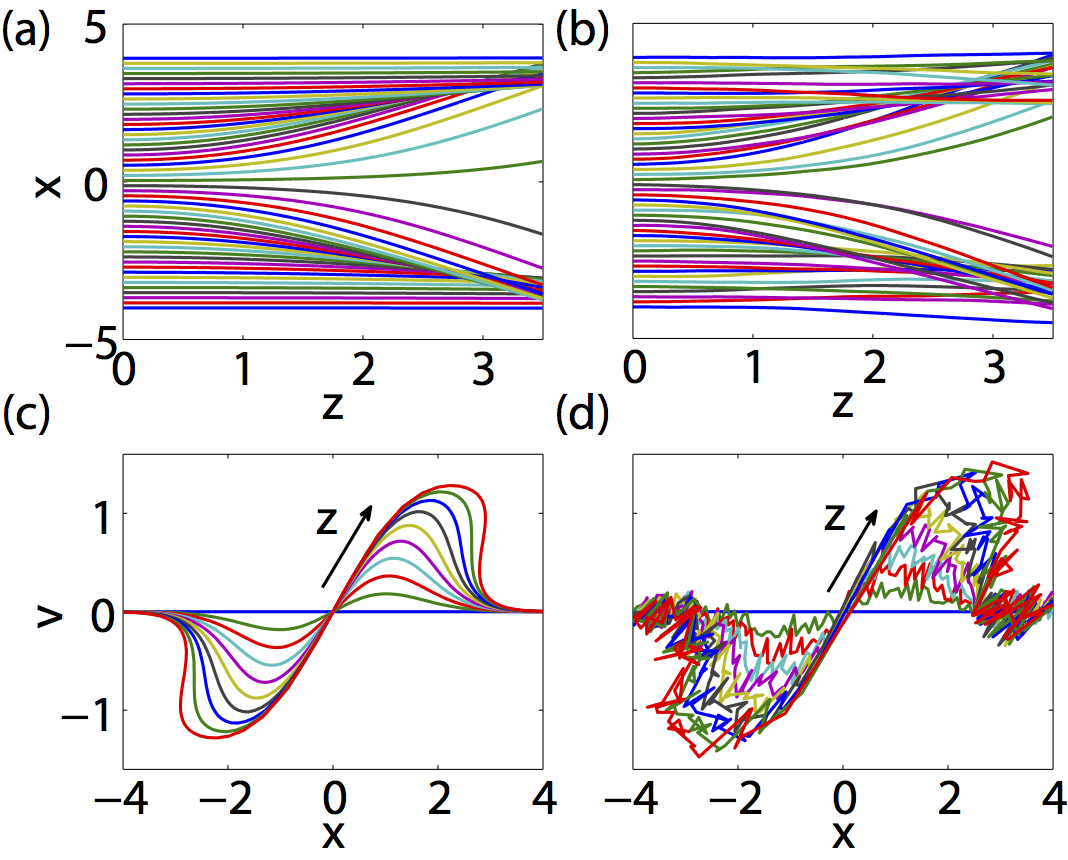}
 \caption{(Color online)
Trajectories of colliding particles forming shock versus $z$: 
(a)  without disorder ($\eta=0$) and, (b), for  $\eta$=0.1;
(c) shock profile in the phase space for $\eta=0$ and, (d),  for $\eta$=0.1
($z$ varies in the range [0,3]).
\label{figth1}
}
\end{figure}

\begin{figure}[h]
\includegraphics[width=8.3cm]{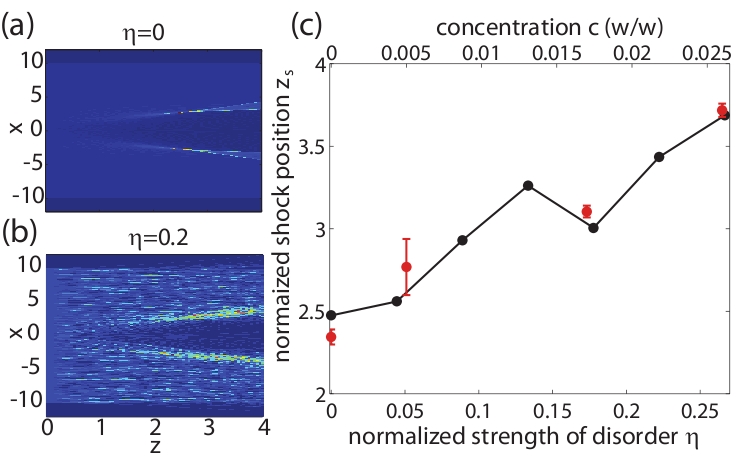}
 \caption{(Color online)
Theoretical histograms of particle positions for $\eta=0$ (ordered case, panel a) and for $\eta=0.2$ (panel b) ;
(c) theoretical normalized shock position $z_s$ versus amount of disorder $\eta$  (black continuous line)
and comparison with the measured $z_s$ Vs concentration $c$ (red squares).
\label{figth2}
}
\end{figure}

In this Letter, we report on the direct experimental evidence of the competition between SWs and disorder, and support our experiments by a theoretical model based on 
the hydrodynamical approximation. We measure the first phase-diagram (where the order parameter is the position of the formation of the shock) for nonlinear waves 
in terms of disorder and nonlinearity, and characterize the scaling laws for the random SWs formation and propagation.
\\\emph{Experiment ---}
We use dispersions of silica spheres of diameter 1$\mu$m in 0.1mM aqueous solutions of RhodamineB displaying a thermal defocusing effect 
due to partial light absorption~\cite{Gho07,Gho09,Gho12,ghofraniha}.
To vary the  degree of disorder several silica  concentrations $c$ are prepared, ranging from 0.005 w/w to 0.03 w/w, in units of weight of silica particles over suspension weight.
A  continuous-wave laser  at wavelength $\lambda$=532~nm is focused  on the input facet of the sample (beam waist  $\simeq 10 \mu$m).
To detect light transmitted at the exit of the samples, the aqueous solutions are put in 1mm$\times$1cm$\times$3cm glass cells
with propagation along the $1$mm  vertical direction (parallel to  gravity) to moderate the effect of heat convection.
Transverse  images of the beam intensity distributions   are collected through an objective  and  recorded  by a 1024$\times$1392 pixel CCD camera.
All measurements are performed after the temperature  
gradient reaches the stationary state and the particle suspensions are
completely homogeneous.
The loss mechanisms in our samples are absorption and scattering. 
The measured loss length (absorption + scattering) is $L\simeq$1.6mm for pure dye solution and  $L\simeq$1.2mm
 for the sample with the highest concentration c (and hence highest losses). These values are obtained by fitting  with exponential decay the beam intensity versus propagation distance Z.
The fact that $L$ is always greater than the position of the shock point $Z_s$ (measured below) allows us to neglect losses at a first approximation in our theory.
In addition, we find that the scattering mean
free path is of the order of millimeters for all the considered
samples. 
\\In Fig.~\ref{figexp1} we show images of the transmitted
beam  (on the $X$-$Y$ transverse plane) for different input laser powers $P$ and various concentrations $c$.
The profiles display post-shock rings
with outer rings being more intense than the inner ones,
as typical for dispersive SWs from Gaussian beams\cite{ghofraniha}.
The  number and the visibility of the oscillations increase with $P$ and decrease with $c$,
evidencing that SWs are sustained by  nonlinearity and inhibited by disorder.
%--------------Figure 1------------------------%
%----------------end figure 1-----------------%
\\We then investigate the onset of the shock along the beam propagation direction: we use 1cm$\times$1cm$\times$3cm glass cells
(propagation along $1$cm), and top images  are collected  through a microscope and recorded by the CCD camera.
The effect of disorder on SWs along beam propagation is
reported in Fig~\ref{figexp2}, where  collected images of the transverse distribution of the beam intensity versus $Z$ for
different input power $P$ and silica concentrations $c$ are shown. 
In accord with Fig.\ref{figexp1}, shock inhibition by disorder in Fig~\ref{figexp2}
is evidenced by the reduction of the beam aperture and the disappearance of the undular bores.
%------------------------Figure 2----------------------%
%------------------------end figure 2----------------------%
\\We identify the point of shock formation $Z_{\textrm{S}}$ as the $Z$ corresponding to the maximum of steepness of the intensity profiles (details  in~\cite{Gho12}). 
We follow this procedure for all images and we report in Fig.~\ref{figexp3}a $Z_{\textrm{S}}$ versus $P$ for different concentrations $c$.
Two effects are evident: 1) for increasing power $P$,  $Z_{\textrm{S}}$  
decreases, corresponding to the speed-up of the shock formation caused by the augmented nonlinearity; 2) for increasing  concentration $c$, $Z_{\textrm{S}}$ increases, 
as disorder delays shock formation up to its total cancellation observed for $c$=0.030~w/w (star symbols).
The plateau at low $P$ indicates that shock is not occurring, as the steepness of the profiles increases with $Z$ but does not have any maximum in the observation window $L_o\simeq 1$~mm.
In this regime  $Z_{\textrm{S}}$ is not the position of the peak of the steepness but that of the highest steepness available in the observable $Z$ range.
We define the value of $P$ at which $Z_{\textrm{S}}$ starts to decrease as the treshold power between shock and non-shock regimes and map the phase diagram in Fig.~\ref{figexp3}b.
%------------------------Figure 3 ----------------------%
%------------------------end 3----------------------%
\\\emph{Theory ---}
In order to theoretically analyze the experimental results,  following previously reported analyses \cite{conti10, ghofraniha}, we use 
the hydrodynamic approximation. We start from the paraxial wave equation for the field complex envelope $A$, in presence of
 disordered local Kerr medium with refractive index perturbation $\Delta n=n_2 I+\Delta n_R(X,Y,Z)$  to the bulk index $n_0$, with $I=|A|^2$
 the optical intensity, $n_2<0$ the Kerr coefficient and $\Delta n_R(X,Y,Z)$ a random perturbation,
\begin{equation}
2i k \frac{\partial A}{\partial Z}+ \nabla^2_{X,Y}A+ 2 k^2 \frac{\Delta n}{n_0} A=0.
\label{eqA}
\end{equation}
In Eq.~(\ref{eqA}), $k=2\pi n_0/\lambda$, and we neglect spatial nonlocality and losses, because they do not qualitatively affect the scenario (as will be reported elsewhere).
The corresponding dimensionless equation for
the normalized field $\psi=A/\sqrt{I_0}$, with $I_0$ the input peak intensity, is
\begin{equation}
i \epsilon \frac{\partial \psi}{\partial z}+\frac{\epsilon^2}{2} \nabla^2_{x,y}\psi- |\psi|^2 \psi + U_R \psi=0\text{,}
\end{equation}
where $(x,y)=(X,Y)/w_0$, $z=Z/L$, $L=\sqrt{L_d L_{nl}}$, $\epsilon=\sqrt{L_{nl}/L_d}$, $L_d=k w_0^2$, $L_{nl}=n_0/(k |n_2| I_0)$,
and $U_R=\Delta n_R/n_2 I_0$ is the ratio between the random index perturbation and the nonlinear one ($w_0\cong 10\mu$m).
%%%%%%%%%%% %%%%%
Because of the huge number of particles  randomly distributed within the optical beam and
the low index contrast (the refractive index is 1.46 for silica and 1.33 for water), $n_R(x, y, z)$ can be taken as a random  dielectric noise
 mainly acting on the phase of the propagating beam; below we show that such an assumption
allows to retrieve theoretical results in quantitative agreement with
experiments.
%%%%%%%%%%%%%%%%%%%%%
In the hydrodynamic limit $\epsilon\rightarrow 0$ ($L_{nl}<<L_d$),
the propagation of the field intensity can be separated by that of the beam phase, 
this results into the  equation of motion for the phase chirp identical to that
of a unitary mass particle (due to the cylindrical symmetry of the system we limit to the $x$-$z$ variables):
\begin{equation}
\frac{d^2 x}{d z^2}=-\frac{dU}{dx}-\frac{dU_R}{dx}=-\frac{dU}{dx}+\eta_R\text{.}
\label{pot1}
\end{equation}
In Eq.~(\ref{pot1}) $U=exp(-x^2/2)$ is the deterministic potential from the nonlinear part due the Gaussian beam profile.
$\eta_R=-dU_R/dx$ is taken as a Langevin force that we assume with Gaussian distribution, such that $\langle \eta_R(z) \eta_R(z') \rangle=\eta^2 \delta(z-z')$, 
with strength of disorder measured by $\eta=\langle (dU_R/dr)^2\rangle^{1/2}\cong<\Delta n_R^2>^{1/2}/|n_2| I_0$,
and the brackets denoting statistical average. 
We stress that in the following we solve eq.(\ref{pot1}) for several values of $x$ taking for each of them an independent realization of the noise $\eta_R(z)$,
this allows to neglect the dependence of $\eta_R$ on $x$.
We stress that for reasons of symmetry the disorder in the two transverse directions are independent.
The simplest and effective theoretical approach is to consider a one-dimensional reduction.
Because of the disorder averaging cylindrical symmetry is preserved, as also experimentally demonstrated.
\\In Fig.\ref{figth1}a,b, we show several of these trajectories resulting from initial uniformly distributed position in the $x$ axis and zero initial velocity $v=dx/dz$,
as obtained by a stochastic Runge-Kutta algorithm \cite{rebecca92}.
Upon propagation the particles collide and, in the absence of disorder, the shock is signaled by
the coalescence of multiple trajectories (Fig. \ref{figth1}a); in the phase space of $v$ and $x$ (Fig.\ref{figth1}c),
these correspond to the folding of the velocity profile into a multivalued function when increasing $z$, which induces the wave-breaking phenomenon.
In the presence of disorder, the particles tend to diffuse, as evident from their trajectories (Fig.\ref{figth1}b) and in the phase space (Fig.\ref{figth1}d);
correspondingly, the propagation distance before their collisions is greater for their random walk and the shock is delayed in the $z-$direction.
\\Figures~\ref{figth2}a,b show the numerically obtained histograms of the particle positions at various propagation distances.
If compared with the ordered case in Fig.\ref{figth2}a, disorder induces a spreading of the particle distribution.
We extract the position of the shock $z_s=z_s(\eta)$ as that approximately corresponding to the
maximum of the histogram (precisely, as the mean value among the positions for which the histogram is above the 90$\%$ of its maximum,
to limit fluctuations).
This allows us to determine $z_s$ for various amounts of disorder $\eta$ (in the ordered case $z_s(0)\cong 2.5$).
Figure~\ref{figth2}c shows $z_s(\eta)$ vs. disorder degree for $10^3$ particles and demonstrates that the shock process is delayed by disorder.

As discussed above, in the absence of disorder, shock appears in the experiments only above a threshold power
(see Fig.\ref{figexp3}a): from a theoretical point of view this threshold arises from the
fact that the hydrodynamic model ($L_{nl}<<L_d$, corresponding to $\epsilon\rightarrow 0$) is valid only at high nonlinearity, hence no shock is expected
at low power. Moreover, in the hydrodynamic limit the position $z=Z \sqrt{|n_2| P/\pi}/w_0^2$ and the shock
position $Z_s$ scales as $1/\sqrt{P}$, as experimentally investigated in \cite{Gho12}.
%--------------Figure 4------------------------%
%----------------end figure 4-----------------%
%--------------Figure 5------------------------%
%----------------end figure 5-----------------%
\noindent Following our theoretical approach, this scaling is maintained in the disordered case and the shock position is delayed,
such that $Z_s(\eta)=z_s(\eta) w_0^2/\sqrt{|n_2|P/\pi}$, $z_s$ is compared 
in Fig.\ref{figth2}c with our experimental results ($|n_2|=2\times10^{-12}$m$^2$W$^{-1}$) revealing quantitative agreement;
discrepancies between experimental and theoretical $z_s$ are ascribed to the several adopted simplifying assumptions in the latter.
\\In addition, from the theory another threshold arises, corresponding to the existence of a critical value for the amount of randomness above which no shock is obtained.
Indeed disorder becomes dominant with respect to nonlinearity when $U_R$ is greater than the deterministic part $U$, such that the hydrodynamic
model in Eq.(\ref{pot1}) is no longer valid. More precisely,  Eq.(\ref{pot1}) holds true as long as $\eta\lessapprox1$,
above this value no shock is expected. As $U$ is of the order of  unity, this corresponds to
$\langle U_R^2 \rangle^{1/2} =\langle \Delta n_R^2 \rangle^{1/2}/ |n_2| I_0
\lessapprox 1$, 
with $\langle \Delta n_R^2 \rangle^{1/2}=c \rho_{H_2 O} (n_{Si O_2}-n_{H_2 O})/\rho_{Si O_2}$, being $n_{Si O_2}$ ($n_{H_2 O}$) 
and $\rho_{Si O_2}$ ($\rho_{H_2 O}$) the refractive index and the density 
of the $Si O_2$ ($H_2 O$), respectively.
That is, no shock is expected when the random index perturbation $\Delta n_R$ 
becomes comparable with the nonlinear perturbation $n_2 I$, such that material fluctuations are so pronounced
that the nonlinear effect is totally masked. 
\noindent In our experiments we have $|n_2|I_0\cong 10^{-3}$,
and we have that the condition $\langle \Delta n_R^2\rangle^{1/2}\cong |n_2| I_0\cong 10^{-3}$  is found for a concentration of   
$c\cong 0.03$~w/w (see figure~\ref{figexp3}c).
In addition, the other boundary line in the phase diagram scales as $\sqrt{P}$ (dot-dashed in figure~\ref{figexp3}c), as retrieved from the condition $Z_s(\eta)<L_o$. \\
\noindent {\it Conclusions ---} By the direct visualization of a laser beam propagating in a liquid random nonlinear system and by imaging  the transmitted light
at the exit of the samples, we show that the gradual increase of disorder hampers shock wave formation, up to its total inhibition.
Such a transition has been quantitatively characterized and results into the first direct measurement of the phase-diagram of nonlinear propagation in terms of disorder
and nonlinearity, as also supported by a theoretical model based on the hydrodynamic approach.
These experiments open the way to further investigations concerning the interplay between disorder and nonlinearity, as the identification of glassy and superfluid phases of light and related phenomena.
\\The research leading to these results has received funding from the European Research Council under the European Community's Seventh
Framework Program (FP7/2007-2013)/ERC grant agreement n. 201766,  from the Italian Ministry of Research (MIUR) through the PRIN project no.2009P3K72Z and from  the Italian Ministry of Education, University
and Research under the Basic Research Investigation
Fund (FIRB/2008) program/CINECA grant code
RBFR08M3P4 and RBFR08E7VA. We thank M. Deen Islam for the technical assistance.

%Bibliografia
%\bibliography{shock}

  \end{document}